\documentclass[aps,prl,reprint,superscriptaddress,amsmath,amssymb]{revtex4-2}

\usepackage{graphicx}
\usepackage{dcolumn}
\usepackage{bm}
\usepackage{xcolor}
\usepackage{bbm}

\usepackage{ulem}

\usepackage{booktabs}

\begin{document}

\preprint{APS/123-QED}

\title{Stress percolation criticality of glass to fluid transition in active cell layers}

\author{Siavash Monfared}
\email{siavash.monfared@nbi.ku.edu}
\affiliation{Division of Engineering and Applied Science, California Institute of Technology, Pasadena, CA 91125, USA.}
\affiliation{Niels Bohr Institute, University of Copenhagen, Denmark.}

\author{Guruswami Ravichandran}
\affiliation{Division of Engineering and Applied Science, California Institute of Technology, Pasadena, CA 91125, USA.}

\author{Jos\'{e} E. Andrade}
\affiliation{Division of Engineering and Applied Science, California Institute of Technology, Pasadena, CA 91125, USA.}

\author{Amin Doostmohammadi}
\email{doostmohammadi@nbi.ku.dk}
\affiliation{Niels Bohr Institute, University of Copenhagen, Denmark.}

\date{\today}

\begin{abstract}
Using three-dimensional representation of confluent cell layers, we map the amorphous solid to fluid phase transition in active cell layers onto the two-dimensional (2D) site percolation universality class. Importantly, we unify two distinct, predominant, pathways associated with this transition; namely (i) cell-cell adhesion and (ii) active traction forces. For each pathway, we independently vary the corresponding control parameter and focus on the emergent mechanical stress patterns as the monolayer transitions from a glassy- to a fluid-like state. Through finite-size scaling analyses, our results lead us to establish the glassy- to fluid-like transition as a critical phenomena in terms of stress development in the cell layer and show that the associated criticality belongs to the 2D site percolation universality class. Our findings offer a fresh perspective on solid (glass-like) to fluid phase transition in active cell layers and can bridge our understanding of glassy behaviors in active matter with potential implications in biological processes such as wound healing, development, and cancer progression. 
\end{abstract}

\maketitle

The transition between solid (glass-like) and fluid phases in cellular systems is of fundamental relevance to a range of biological processes, including cancer metastasis~\cite{oswald2017jamming,Grosser_2021,Blauth_2021}, wound healing~\cite{Kim_2013,Tetley_2019,Jain_2020} and tissue morphogenesis~\cite{mongera2018fluid,Petridou_2018,Petridou_2019}. Over the past decade, this transition is revealed to be governed by collective cell organization in biological tissues~\cite{Farhadifar_2007,Trepat_2009,Angelini2010,Nnetu_2012,Sch_tz_2013}, spurring a plethora of research  on this topic ~\cite{Petridou_2019,Atia_2021,Lawson_Keister_2021,Hannezo_2022}. 
To understand the solid-fluid transition in biological systems, the concept of jamming has been applied to inherently out-of-equilibrium living cells~\cite{Garcia2015,Bi2015,Atia2018} where jamming-unjamming is used to refer to the transition between solid-like glassy and fluid-like states. This is despite the geometrical roots of jamming transition and its correspondence to the zero temperature and zero activity limit of a glass transition~\cite{Liu1998,Cates1998,Parisi_2010,berthier2019glassy}. In this vein, non-equilibrium glass transition in active systems share many features with equilibrium glass transition~\cite{Berthier_2013} and an effective thermodynamic equilibrium may prevail for moderate activity driving the system out-of-equilibrium~\cite{Fodor_2016}.  

Over the past decade, the glass-like to fluid transition in living cells is broadly understood as a geometric jamming transition characterized by a critical shape index $q=\mathcal{P}/\sqrt{\mathcal{A}}$, where $\mathcal{P}$ and $\mathcal{A}$ are cell perimeter and area respectively. This is primarily based on the application of two-dimensional vertex models \cite{Park2015,Bi2015}, a voronoi based variant endowed with motility \cite{Bi2016} and experiments on a range of epithelial systems \cite{Atia2018} underlining the role of geometrical constraints. More recent studies highlight the role of two types of percolations based on (i) cell connectivity \cite{Petridou_2021} and (ii) edge tension network \cite{Li_2019,Fuhs_2022}. In particular, experiments on a zebrafish blastoderm \cite{Petridou_2018}, a non-confluent cell system with synchronous cell divisions, show a solid-fluid transition at the onset of zebrafish morphogenesis. This transition is linked to a rigidity percolation based on cell connectivity~\cite{Petridou_2021}. On the other hand, application of a vertex model to heterogeneous cell layers \cite{Li_2019} and experiments on primary tumour explants \cite{Fuhs_2022} show rigidity percolation based on edge tension network gives rise to finite shear modulus in tumor explants consisting of heterogeneous mixture of soft and stiff cancer cells. Notwithstanding these seminal and important contributions, the universality of the transition between active solid (glass-like) and fluid phases in cellular systems and its broader applicability is yet to be established~\cite{Sussman_2018,Sussman_2018a,mongera2018fluid,Czajkowski_2019,Saraswathibhatla2020,kim2021embryonic,Petridou_2021,Hopkins_2022}. 
\begin{figure}[t]
\includegraphics[trim={0.8cm 4.4cm 1.0cm 3.cm},clip,width=0.5\textwidth]{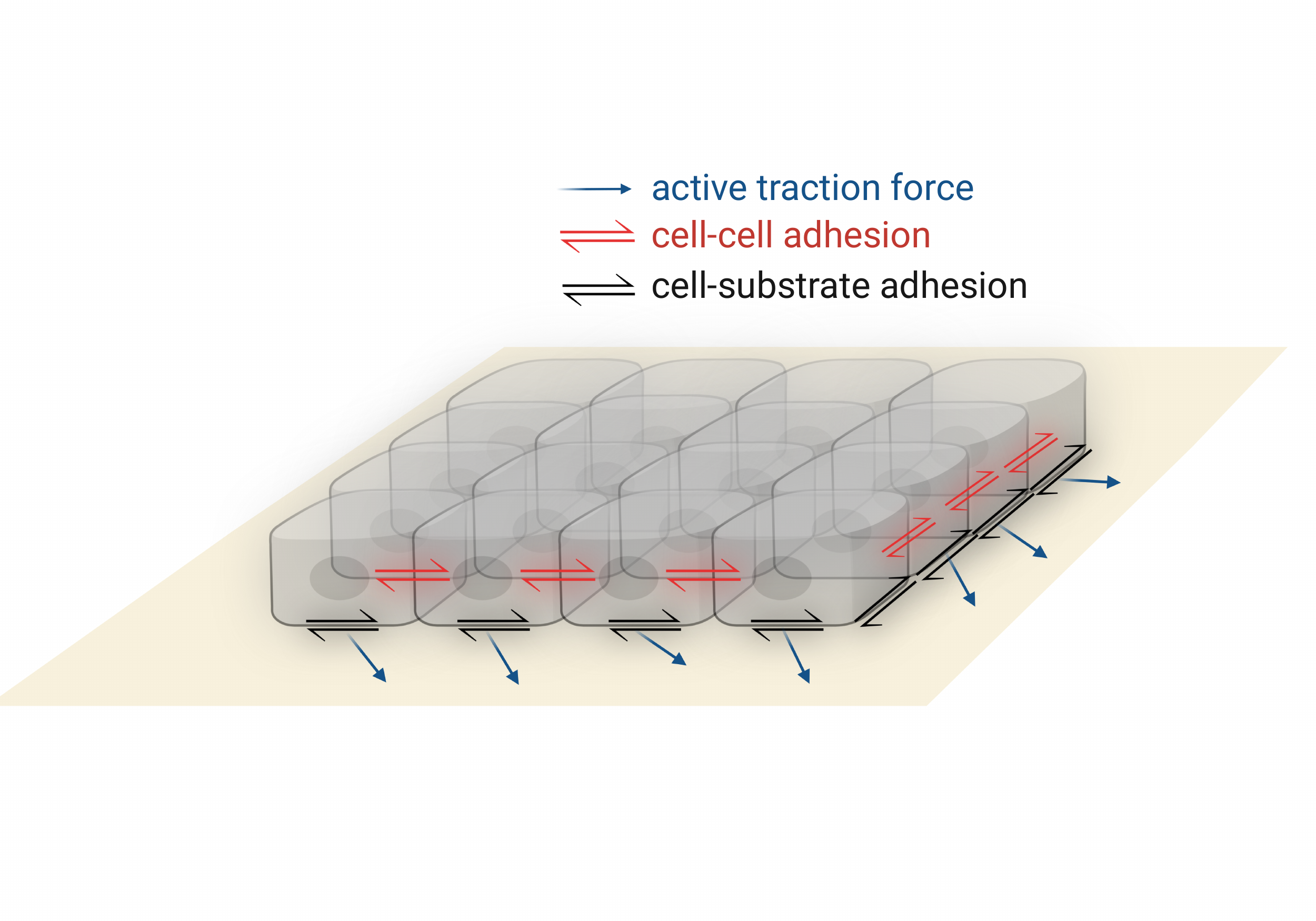}
\caption{Schematic of a confluent layer displaying cell-cell and cell-substrate adhesions and in-plane active traction forces acting on the plane of the substrate (illustrated as yellow surface). Created with BioRender.com}
\label{fig_schematics}
\end{figure}
Therefore, despite the immense significance of transition from glassy- to fluid-like state in cell collectives in various biological processes, the nature of transition remains elusive.

\begin{figure*}[t]
\includegraphics[trim={0.cm 0.cm 0.cm 0.cm},clip,width=\linewidth]{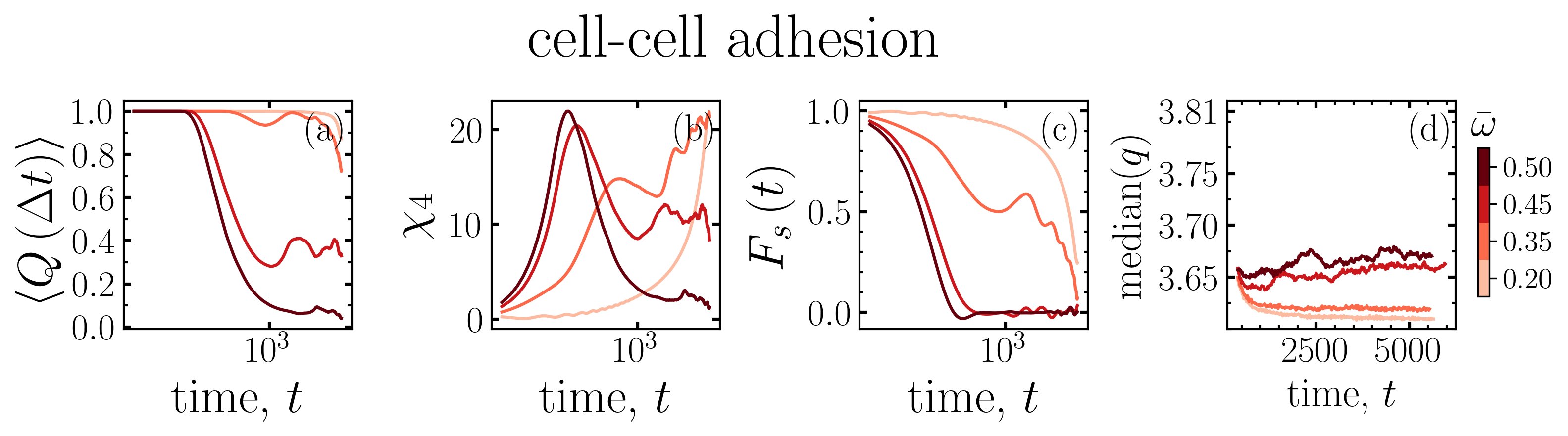}
\includegraphics[trim={0.cm 0.cm 0.cm 0.cm},clip,width=\linewidth]{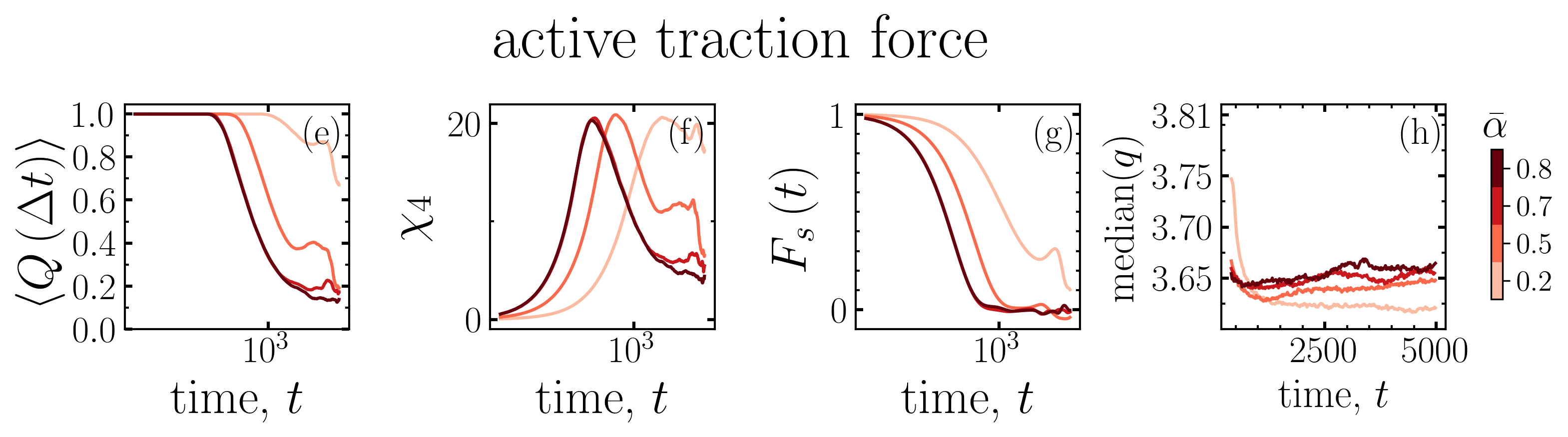}
\caption{Quantitative characterization of cell mobility in terms of overlap function (a,e), four-point susceptibility (b,f) and self-scattering function (c,g) as cell-cell adhesion (a,b,c) and active traction force strength (e,f,g) are increased. The temporal evolution of the median of shape parameter, $q$ with increasing $\bar\omega$ (d) and $\bar\alpha$ (h). }
\label{fig_mobility}
\end{figure*}

Here, we characterize the criticality of the active glass-to-fluid transition in cell layers. Importantly, we establish this through two distinct paths by varying, independently, cell-cell adhesion and active traction forces as control parameters and recover the 2D site percolation critical exponents for isotropic stress clusters via finite-size scaling analyses.

We build on a recently developed three-dimensional phase-field model for cell layers, which allows for independent variation of cell-cell and cell-substrate interactions~\cite{Monfared2022}.
We consider a cellular monolayer consisting of $N=400$ cells on a substrate with its surface normal $\vec{e}_{n}\left(=\vec{e}_{z}\right)=\vec{e}_{x}\times\vec{e}_{y}$ and periodic boundaries in both $\vec{e}_{x}$ and $\vec{e}_{y}$, where $\left(\vec{e}_{x},\vec{e}_{y},\vec{e}_{z}\right)$ constitute the global orthonormal basis (see Fig. \ref{fig_schematics}). Each cell $i$ is modeled as an active deformable droplet in three-dimensions using a phase-field, $\phi_{i}=\phi_{i}\left(\vec{x}\right)$ and initialized with radius $R_{0}$. The interior and exterior of cell $i$ corresponds to $\phi_{i}=1$ and $\phi_{i}=0$, respectively, with a diffuse interface of length $\lambda$ connecting the two regions and the midpoint, $\phi_{i}=0.5$, delineating the cell boundary. 

This approach resolves the cellular interfaces and provides access to intercellular forces. The dynamics of the phase field $\phi_{i}$ is evolved through: 
\begin{eqnarray}
\label{eq2}
\partial_{t}\phi_{i}+\vec{v}_{i}\cdot\vec{\nabla}\phi_{i}=-\frac{\delta\mathcal{F}}{\delta\phi_{i}},\qquad i=1,...,N, 
\end{eqnarray}
where $\vec{v}_{i}$ is the velocity of cell $i$ and $\mathcal{F}[\phi]$ is the three-dimensional free energy functional that stabilizes cell interface and accounts for cell mechanical properties including cell stiffness ($E$) and compressibility ($\mu$), and puts a soft constraint on the cell volume~\cite{Palmieri2015,aranson2016physical,camley2017physical,Mueller2019} around $V_{0}=(4/3)\pi R_{0}^{3}$. Additionally, the free energy comprises gradient contributions ($\vec{\nabla}\phi$) that account for, and distinguish between, cell-cell ($\omega_{cc}$) and cell-substrate ($\omega_{cw}$) adhesions, as introduced recently~\cite{Monfared2022}:
\begin{eqnarray}
\label{eq1}
\mathcal{F} &=& \sum_{i}^{N}\frac{E}{\lambda^2}\int d\vec{x}\{4\phi_{i}^{2}\left(1-\phi_{i}\right)^{2}+\lambda^{2}\left(\vec{\nabla}\phi_{i}\right)^{2}\}\notag \\ 
 &+& \sum_{i}^{N}\mu\left(1-\frac{1}{V_{0}}\int d\vec{x}\phi_{i}^{2}\right)^{2}+\sum_{i}^{N}\sum_{j\neq i}\frac{\kappa_{cc}}{\lambda}\int d\vec{x}\phi_{i}^{2}\phi_{j}^{2}\notag\\
 &+& \sum_{i}^{N}\sum_{j\neq i}\frac{\omega_{\text{cc}}}{\lambda^{2}}\int d\vec{x}\vec{\nabla}\phi_{i}\cdot\vec{\nabla}\phi_{j}+\sum_{i}^{N}\frac{\kappa_{cw}}{\lambda}\int d\vec{x}\phi_{i}^{2}\phi_{w}^{2}\notag\\
 &+& \sum_{i}^{N}\frac{\omega_{\text{cw}}}{\lambda^{2}}\int d\vec{x}\vec{\nabla}\phi_{i}\cdot\vec{\nabla}\phi_{w},
\end{eqnarray}
where $\kappa$ captures repulsion between cell-cell (subscript $cc$) and cell-substrate (subscript $cw$) and $\phi_{w}$ denotes a static phase-field representing the substrate. To resolve the forces generated at the cellular interfaces, we consider the following over-damped dynamics for cells: 
\begin{equation}
\label{eq3}
\vec{T}_{i} = \zeta \vec{v}_{i} - \vec{F}^{\text{act}}_{i} = \int d \vec{x} \phi_{i} \vec{\nabla} \cdot \left(\sum_{i}-\left(\delta\mathcal{F}/\delta\phi_{i}\right)\right)\bm{1}
\end{equation}


where $\vec{T}_{i}$ denotes traction~\cite{Trepat_2009,Trepat2014,Saw2017} and contains both active and passive contributions, $\zeta$ is substrate friction and $\vec{F}_{i}^{\text{act}}=\alpha\hat{F}_{i}^{\text{pol}}$ represents self-propulsion force directed by cell polarity, constantly pushing the system out-of-equilibrium with $\alpha$ characterizing the strength of the self-propulsion force, and $\hat{F}_{i}^{\text{pol}}$ a unit vector corresponding to the polarity direction for each cell. The dynamics of cell polarity is introduced based on contact inhibition of locomotion~\citep{Abercrombie_1954,Abercrombie_1979} by aligning the polarity of the cell to the direction of the total interaction force acting on the cell~\citep{Smeets2016}, and is chosen specifically based on recent experimental observations~\cite{Peyret2019} (see {\it Materials and Methods}). Throughout this study we use the initial cell radius $R_{0}$ as the characteristic length scale, the polarity alignment time $\tau$ as the characteristic time scale, and the cell stiffness $E$ as the characteristic force scale to report the results in dimensionless units (see {\it Materials and Methods}).

In passive, athermal systems, jamming transition is intimately related to the packing fraction (density) of that system. However, for living cells at confluence, the collective behavior can change from a fluid-like to a solid (glass-like) behavior at a constant density. In recent studies, density is also shown to have a second order effect on fluid-solid transition in cellular systems \cite{Garcia2015}. With this in mind, we keep the cell density constant in our simulations, i.e. we do not allow cell proliferation nor cell extrusion, while ensuring confluency. For each considered pathway, we perform large scale simulations by incrementally increasing the dimensionless cell-cell to cell-substrate adhesion ratio $\bar{\omega}=\omega_{cc}/\omega_{cw}\in[0.1,0.5]$ for the first, and the dimensionless traction force $\bar{\alpha}=\alpha \tau / \zeta R_{0}$, $\bar{\alpha}\in[0,0.8]$, which characterizes the ratio of self-propulsion to the friction with the substrate, for the second pathway. For each case, three distinct realizations and $n_{\text{sim}}=5000$ time steps are considered. As we show next, the considered ranges for $\bar{\omega}$ and $\bar{\alpha}$ captures the glassy- to fluid-like state of cells. Other simulation parameters are fixed and outlined in the {\it Materials and Methods}.

We begin by establishing what constitutes as a glassy state, which is necessary to identify the onset of glassy- to fluid-like phase transition. To this end, we characterize cell displacement and cooperative cell motion as a function of the two control parameters, i.e. dimensionless cell-cell adhesion $\bar{\omega}$, and strength of traction force $\bar{\alpha}$ (Fig. \ref{fig_mobility}). 

First we quantify the overlap function, the fractional change of cellular position in a given time increment $\Delta t$, $Q\left(\Delta t\right)=\left(1/N\right)\sum_{i=1}^{N}w_{i}$, where $w=1$ if $|\vec{x}_{i}\left(t+\Delta t\right)-\vec{x}_{i}\left(t\right)|<R_{0}$ and $w=0$, otherwise. As the strength of cell-cell adhesion $\bar\omega$, and the self-propulsion force $\bar{\alpha}$ increase, the overlap function falls below $0.2$, reflecting the high mobility of the cells and potentially exchanging neighbors in a fluid-like state (Fig.\ref{fig_mobility}a,e). 

\begin{figure*}[t]
\includegraphics[trim={0.cm 0.cm 0.cm 0.cm},clip,width=0.49\textwidth]{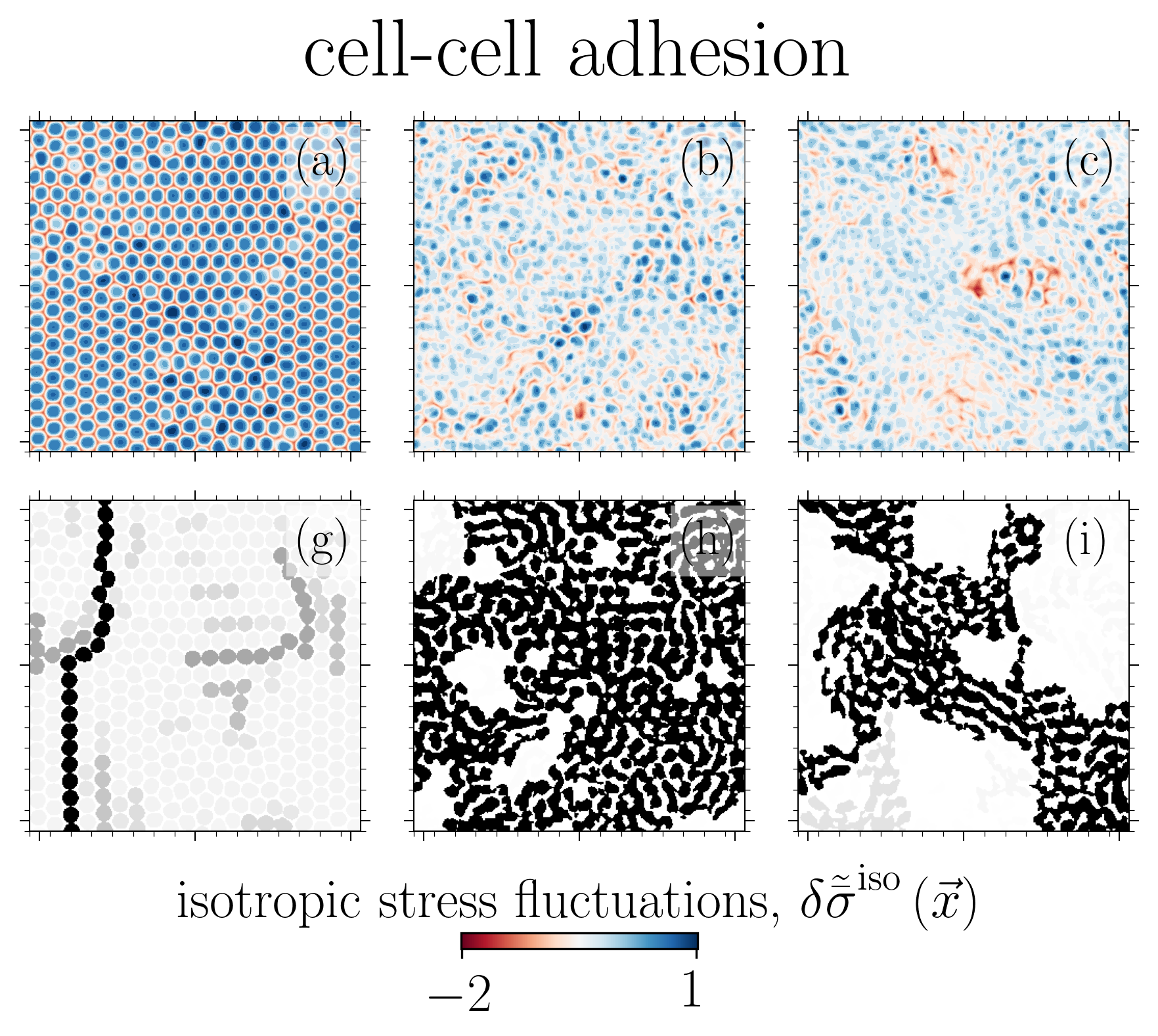}
\includegraphics[trim={0.cm 0.cm 0.cm 0.cm},clip,width=0.49\textwidth]{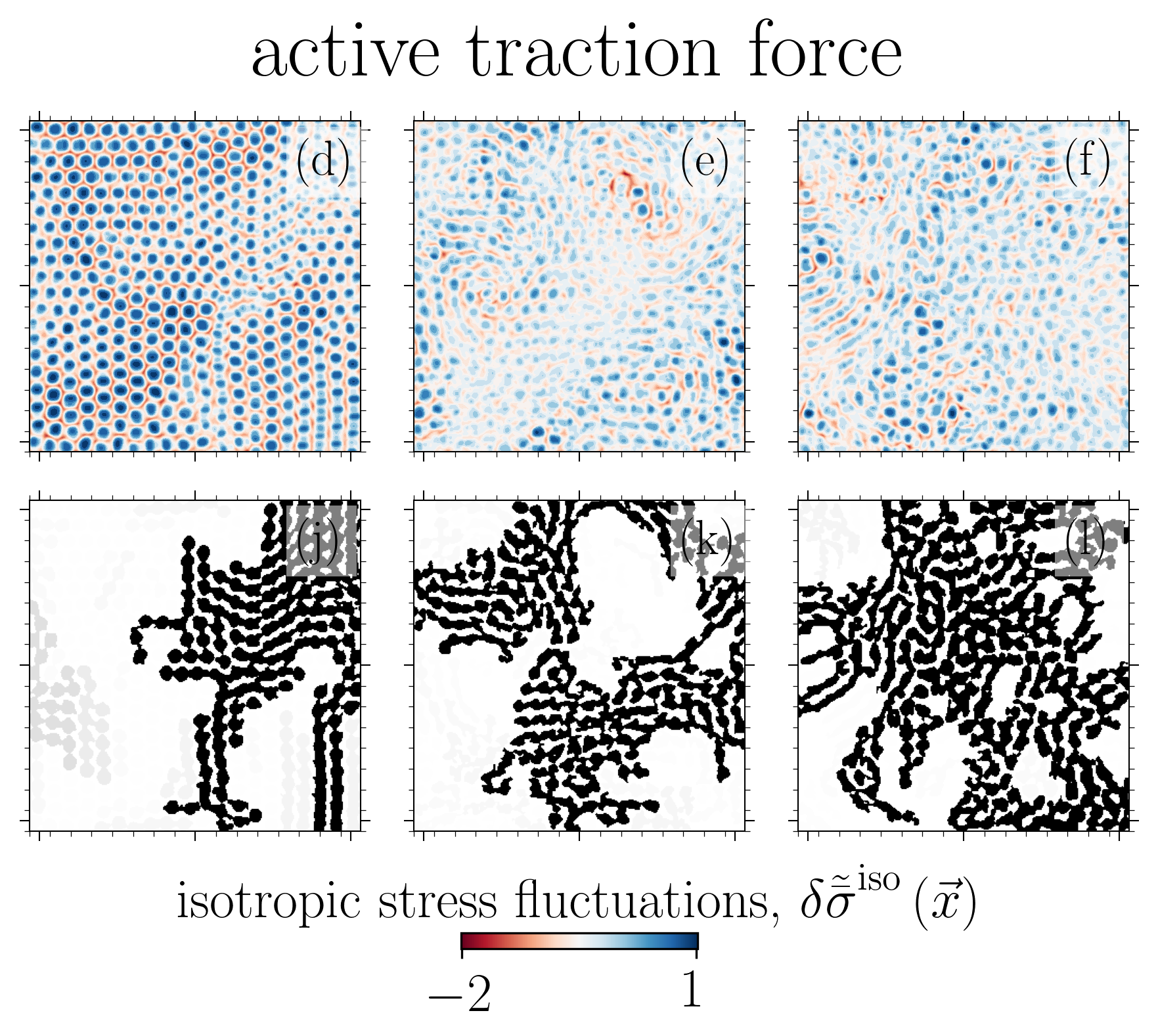}
\caption{Percolation of isotropic stress within the cell layer. The fluctuations in normalized temporal average of the isotropic stress field, $\delta\tilde{\bar{\sigma}}^{\text{iso}}=\tilde{\bar{\sigma}}^{\text{iso}}-\langle\tilde{\bar{\sigma}}^{\text{iso}}\rangle$, $\tilde{\bar{\sigma}}^{\text{iso}}\left(\vec{x}\right)=\bar\sigma^{\text{iso}}/\sigma^{\text{iso}}_{\text{max}}$ where $\sigma^{\text{iso}}=(1/3)\text{tr}\bm\sigma$ for increasing dimensionless cell-cell adhesion (a) $\bar\omega=0.2$, (b) $\bar\omega=0.45$ and (c) $\bar\omega=0.5$. The largest spanning cluster (in Black) corresponding to  $\tilde{\bar{\sigma}}^{\text{iso}}\left(\vec{x}\right)$ for (g) $\bar\omega=0.2$ percolating at $\rho=0.743$, (h) $\bar\omega=0.45$ percolating at $\rho=0.647$, and (i) $\bar\omega=0.5$ percolating at $\rho=0.525$ and dimensionless active traction strength $\bar\alpha$ for
(d) $\bar\alpha=0.2$, (e) $\bar\alpha=0.7$ and (f) $\bar\alpha=0.8$. The largest spanning cluster (in Black) corresponding to  $\tilde{\bar{\sigma}}^{\text{iso}}\left(\vec{x}\right)$ for (j) $\bar\alpha=0.2$ percolating at $\rho=0.622$, (k) $\bar\alpha=0.7$ percolating at $\rho=0.540$, and (l) $\bar\alpha=0.8$ percolating at $\rho=0.599.$}
\label{fig_stress}
\end{figure*}

To quantify the cooperative motions of cells, we use the overlap function, $Q\left(\Delta t\right)$, to compute the four-point susceptibility, $\chi_{4}=N[\langle Q\left(\Delta t\right)^{2}\rangle-\langle Q\left(\Delta t\right)\rangle^{2}]$. When cells move cooperatively, $\chi_{4}$ exhibits a clear peak with its position and magnitude corresponding to pack lifetime and pack size, approximately~\cite{Angelini2011}. For a solid (glass-like) state, typically the swirl lifetime grows and exceeds the temporal window of observation, in our case that is the total simulation time. For both the strong cell-cell adhesion, $\bar{\omega}=0.5$, and self-propulsion force strength $\bar\alpha=0.8$, a fluid-like state is observed. This is evident from the overlap function $\langle Q\left(\Delta t\right) \rangle$ indicating the formation of a pack of $\sim 20$ cells with a lifetime of $\sim 400$ time steps, in dimensionless units (Fig. \ref{fig_mobility}b,f). For the adhesion or self-propulsion strengths lower than this critical value, peaks in $\chi_{4}$ are observed, with the pack size increasing as the adhesion or self-propulsion force strength is increased.

To complete the characterization of the glassy dynamics we also quantify the self-scattering function, $F_{s}\left(k,t\right)=\langle e^{i\vec{k}\cdot\Delta\vec{r}\left(t\right)}\rangle$ where $|\vec{k}|=\pi/r_{0}$, and $r_{0}$ is the position of the first peak in the pair-correlation function. Consistent with the characterization of the overlap function and four-point susceptibility, for the highest value of cell-cell adhesion, $F_{s}$ approaches zero indicative of a fluid-like state (Fig. \ref{fig_mobility}c). For $\bar\omega\ll 0.5$, $F_{s}$ is around one suggesting a glass state. Similar behavior is observed for the strength of self-propulsion force $\bar\alpha\ll 0.8$ (Fig. \ref{fig_mobility}g). 
Together, the results of the measurements of the overlap function, four-point susceptibility, and the self-scattering function confirm that upon increasing the ratio of cell-cell adhesion to cell-substrate adhesion, or independently, increasing the self-propulsion force, the active cell layer continuously transitions from a solid (glassy) state to a fluid-like state. Particularly, based on the presented analyses of cell cooperative motions, this transition takes place at $\bar\omega=0.5$ and $\bar\alpha=0.8$.

The results discussed so far, indicating a glass-to-fluid transition upon increasing cell-cell adhesion or self-propulsion force, is consistent with the previous results observed in experiments and vertex models of cell layers~\cite{Garcia2015,Atia2018,Bi2016}. 
To this end, our point of departure from the previous characterization of the active glass-to-fluid transition in cell layers is that by accounting for both cell-cell and cell-substrate interactions we do not observe any geometric signatures for jamming transition associated with the shape order parameter. This is evident from Fig.~\ref{fig_mobility}d,h where we compute the projected area and the perimeter for each cell and every simulation time steps and compute the cell shape index, $q$. At all times after the initialization, for both glassy- and fluid-like systems, the median for the shape index, $\bar{q}$ remains under $\bar{q}=3.75$, clearly deviating from the value of $\bar{q}\approxeq 3.81$ that has been suggested as the structural order parameter for solid-to-fluid jamming transition in cell layers~\cite{Bi2015}. Similar deviation has been observed in experiments on zebrafish embryo and the associated active vertex model that allows for finite extracellular spaces between the cells~\cite{kim2021embryonic}. Moreover, detailed analysis of the voronoi based model \cite{Bi2016} has find no geometric jamming transition at the zero temperature limit \cite{Sussman_2018}. Additionally, recent experiments on Madin-Darby canine kidney (MDCK) cells do not show such geometric criticality and through various molecular perturbations establish that the change in adhesion strength has no effect on the cell perimeter, instead demonstrating that the traction forces play a dominant role in glass to fluid phase transition in the epithelial tissues~\cite{Saraswathibhatla2020}.

\begin{figure*}[t]
\includegraphics[trim={0.cm 0.cm 0.cm 0.cm},clip,width=\textwidth]{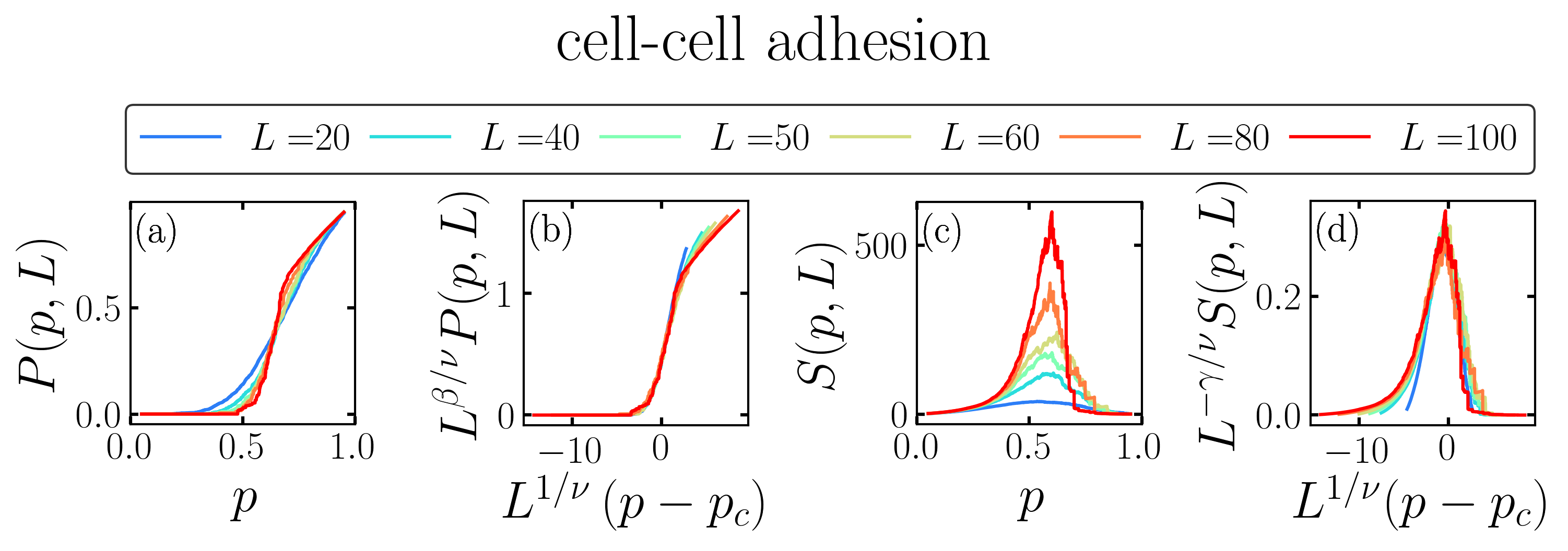}
\includegraphics[trim={0.cm 0.cm 0.cm 0.cm},clip,width=\textwidth]{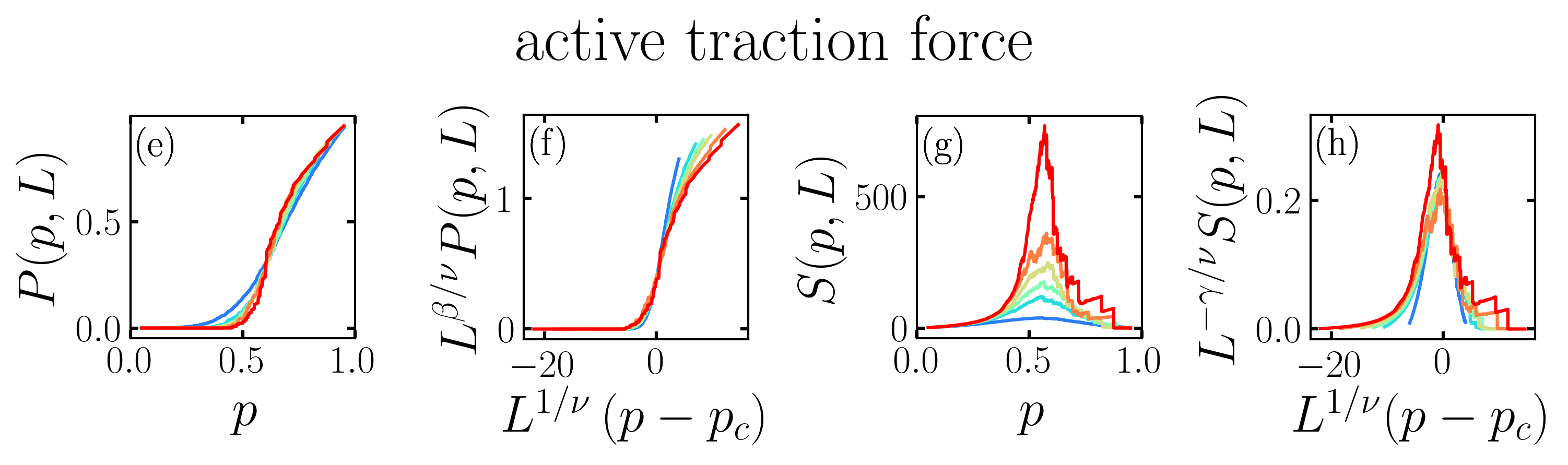}
\caption{Finite-size scaling analyses of isotropic stress criticality. The density of spanning cluster $P(p,L)$ (a,e) and the average cluster size $S(p,L)$ at different length scales $L$ and their collapse after performing a finite size scaling analysis for $P(p,L)$ (b,f) and $S(p,L)$ (c,h). First row (a-d) corresponds to increasing dimensionless cell-cell adhesion $\bar\omega=\omega_{cc}/\omega_{cw}$ and the second row (e-h) corresponds to increasing strength of active traction forces $\bar\alpha$.}
\label{fig_fss}
\end{figure*}

Rather than geometric considerations, to understand the glass transition in the considered confluent active cell layers, we turn to the emergent mechanical stress fields. To this end, we compute a coarse-grained stress field \cite{Christoffersen1981}:
\begin{equation}\label{eq_stress}
\sigma_{ij}=\frac{1}{2V_{\text{cg}}}\sum_{m\in V_{\text{cg}}}\left(\vec{T}_{i}\left(\vec{x}_{m}\right)\otimes\vec{e}_{j}^{n}+\vec{T}_{j}\left(\vec{x}_{m}\right)\otimes\vec{e}_{i}^{n}\right)
\end{equation}
where $\vec{x}_{0}$ represents the center of the coarse-grained volume, $V_{\text{cg}}=\ell_{cg}^{3}$, corresponding to coarse-grained length $\ell_{cg}=a_{0}$ where $a_{0}$ is the unit grid length, 
and unit vector $\vec{e}^{n}_{i}=\left(\vec{x}_{0}-\vec{x}_{m}\right)/|\vec{x}_{0}-\vec{x}_{m}|$. Given the definition of $\vec{T}_{i}$ in \eqref{eq3}, the coarse grained stress field, \eqref{eq_stress}, contains contributions from both active and passive forces. To probe for a possible mechanical criticality at the active glass-to-fluid transition, we then measure the temporally averaged isotropic stress field $\tilde{\bar{\sigma}}^{\text{iso}}\left(\vec{x}\right)=\bar{\sigma}^{\text{iso}}\left(\vec{x}\right)/\bar{\sigma}^{\text{iso}}_{\text{max}}\left(\vec{x}\right)$, where $\bar{\sigma}^{\text{iso}}\left(\vec{x}\right)=\left(1/n_{\text{sim}}\right)\sum^{n_{\text{sim}}}_{t}\sigma^{\text{iso}}\left(\vec{x},t\right)$, with $n_{\text{sim}}$ number of simulation time steps and $\sigma^{\text{iso}}\left(\vec{x},t\right)=\left(1/3\right)\text{tr}\bm\sigma\left(\vec{x},t\right)$, provides a measure of the amount of expansion ($\sigma^{\text{iso}} > 0$) and compression ($\sigma^{\text{iso}} < 0$) in the cell layer. 
Visual inspection of the fluctuations in isotropic stress fields and the associated patterns shows that for both adhesion parameter $\bar\omega$ (Figs. \ref{fig_stress}a-c) and traction force $\bar\alpha$ (Figs. \ref{fig_stress}d-f), as the control parameter is increased and the system approaches a fluid-like state, a more disordered isotropic stress field emerges, suggesting a possible percolation of isotropic stress in the system. Figs. \ref{fig_stress}g-l display the spanning cluster - a cluster of connected sites that spans two opposing sides of the system's boundary - in Black, for each pathway as $\bar\omega$ or $\bar\alpha$ is increased. To quantify this emergent build up and the possible links to percolation universality classes, we perform a finite-size scaling analysis~\cite{Fisher_1967,Fisher_1972} for each path. We begin with quantifying the density of the spanning cluster, $P\left(p,L\right)$, which is the probability of a site belonging to a spanning cluster as a function of the occupation probability, $p$, and system size, $L$. The occupation probability, $p\in[0,1]$, corresponds to sites where $\tilde{\bar{\sigma}}^{\text{iso}}\left(\vec{x}\right)$ is greater than $\left(1-p\right)\times 100$-th percentile of the isotropic stress distribution. In the thermodynamic limit, i.e. $L\rightarrow\infty$, we expect the following power-law scaling near the percolation probability, $p_{c}$ for the density of spanning cluster $P\left(p\right)$ characterized by critical exponent $\beta$:
\begin{equation}
P\left(p\right)\sim\left(p-p_{c}\right)^{\beta}.
\end{equation}
Furthermore, we quantify the average cluster size, $S\left(p\right)=\langle s\rangle$ where $s$ is size of a cluster. Near the percolation probability, the power-law scaling of $S\left(p\right)$ quantified by critical exponent $\gamma$ reads:
\begin{equation}
S\left(p\right)\sim|p-p_{c}|^{-\gamma},
\end{equation}
and critical exponent $\nu$ which describes the power-law scaling of correlation length $\xi$, an average distance between two sites in the same cluster, is expected to follow:
\begin{equation}\label{eq_nu}
\xi\sim|p-p_{c}|^{-\nu}.
\end{equation}

\begin{table*}
\centering
\caption{\label{tab_fss}Critical exponents obtained from finite size scaling analysis.}
\begin{tabular}{lcccc}
{} & $p_{c}$ & $\gamma$ & $\beta$ & $\nu$\\
\midrule
2D site percolation~\cite{Stauffer_2018} & $0.5927$ & $2.388$ & $0.1388$ & $1.333$\\
This study ($\bar\omega$) & $0.612\pm0.004$ & $2.305\pm0.146$ & $0.176\pm0.049$ & $1.422\pm0.051$\\
This study ($\bar\alpha$) & $0.590\pm0.007$ & $2.103\pm0.423$ & $0.134\pm0.009$ & $1.243\pm0.198$\\
\bottomrule
\end{tabular}
\end{table*}
Both the density of the spanning cluster, $P\left(p,L\right)$ and the average cluster size $S\left(p,L\right)$ are displayed in Fig. \ref{fig_fss} at the onset of the transition, i.e. $\bar\omega=0.45<0.5$ (Fig. \ref{fig_fss}a,c) and $\bar\alpha=0.7<0.8$ (Fig. \ref{fig_fss}e,g) for various length scales, $L$. Additionally, Fig. \ref{fig_fss}b,d,f,h show the collapse of $P\left(p,L\right)$ and $S\left(p,L\right)$ when scaled with critical exponents obtained from the finite-size scaling analysis, accounting for a diverging correlation length $\xi$ near the $p_{c}$, i.e. \eqref{eq_nu}. Remarkably, the critical exponents corresponding to the scaling of temporally averaged isotropic stress field at the onset of the glass-to-fluid transition in the active cell layer (Tab. \ref{tab_fss}) lead to a reasonable collapse and are in close agreement with those from the 2D site-percolation universality class \cite{Stauffer_2018} (see supporting information for more details regarding finite-size scaling analysis). Importantly, the agreement with the critical exponents for 2D site-percolation universality class are obtained for two independent pathways (i.e., cell-cell adhesion and active traction force) of driving the system through glass-to-fluid transition, further reinforcing the universality of the isotropic stress percolation at the onset of active glass-to-fluid transition in model cell layers. Moreover, a recent study suggest that the microscopic details of self-propulsion does not affect the glassy dynamics of active systems \cite{Debets_2022}. This implies that our presented analyses would hold irrespective of how we introduce active self-propulsion forces, $\vec{F}^{\text{act}}_{i}=\alpha\hat{F}^{\text{pol}}_{i}$ (see {\it Materials and Methods} for details regarding the dynamics of cell polarity). 

Our findings contribute to the mounting evidence that suggest the existence of critical phenomena and universality classes in diverse range of active and biological systems including a first-order phase transition associated with the cell plasticity~\cite{font2018topography,la2020phase},
\textit{epithelial-to-mesenchymal} (EMT) during tissue spreading as a wetting transition~\cite{Douezan_2011,Prez_Gonzalez_2018}, 
and viscosity changes due to rigidity percolation in fly embryo~\cite{Petridou_2021}.

Our results can also help guide the conversation on the origins of density-independent solid-to-fluid phase transition in biological systems. Currently, the consensus on dominant mechanisms responsible for such a transition are: (i) shape order parameter, (ii) rigidity percolation based on cell-cell connectivity modulated by cell-cell adhesion and (iii) percolation of mechanical tension. All of these mechanisms are implicitly manifested in stress fields and thus our findings, including the mapping to the 2D site percolation universality class, can potentially unify previously reported experimental and theoretical observations. To this end, our work opens new possibilities to further investigate the nature of glass-to-fluid transition in active systems using the tools of non-equilibrium statistical mechanics and to explain transitionary behaviours in biological systems. Moreover, the reported findings call for development of new theoretical formulations to explain active glass-to-fluid transition based on emerging stress patterns and the criticality of the isotropic stress percolation. Simultaneously, direct measurement of traction forces and mechanical stress fields produced by multicellular layers is currently experimentally accessible~\cite{Ladoux2009,Trepat2014} and can be used to probe the proposed universality in cellular monolayers in experiments.

Lastly, it would be interesting to further study any possible connections between these results and those for glass transition and jamming in passive, weakly connected amorphous solids and granular materials. Particularly intriguing is the association of jamming in passive systems with the emergence of rigidity \cite{O_Hern_2003,Wyart_2005} and a variety of percolation models \cite{Toninelli_2006,Schwarz_2006,Jeng_2010}. 

\subsection*{Simulation parameters}
We consider a cellular monolayer consisting of $N=400$ cells on a substrate with its surface normal $\vec{e}_{n}\left(=\vec{e}_{z}\right)=\vec{e}_{x}\times\vec{e}_{y}$ and periodic boundaries in both $\vec{e}_{x}$ and $\vec{e}_{y}$, where $\left(\vec{e}_{x},\vec{e}_{y},\vec{e}_{z}\right)$ constitute the global orthonormal basis. Cells are initiated on a two-dimensional simple cubic lattice and inside a cuboid of size $L_{x}=L_{y}=320$, $L_{z}=64$, grid size $a_{0}=1$ and with radius $R_{0}=8$. Simulations are run for $n_{\text{sim}}=5000$ time steps.
We perform large scale simulations with a focus on the interplay of cell-cell and cell-substrate adhesion strengths on collective cell migration and its impact on cell expulsion from the monolayer. Following \citep{Mueller2019}, the space and time discretization in our simulations are based on the average radius of MDCK cells, $\sim 5\mu m$, velocity $\sim 20\mu m/h$ and average pressure of $\sim 100$Pa, measured experimentally in MDCK monolayers \citep{Saw2017}, corresponding to $\Delta x\sim0.5\mu m$, $\Delta t\sim 0.1s$ and $\Delta F\sim 1.5$nN for force. The simulation parameters $\kappa_{\text{cc}}=0.5$, $\kappa_{\text{cw}}=0.15$, $\zeta=1$, $E=0.024$, $\mu=45$, $D_{r}=0.01$ and $\tau=200$, are chosen in the range that was previously shown successful for comparison of phase-field model with experimental measurements of sustained oscillations~\citep{Peyret2019} and characterization of flows and topological defects in eputhelial cell monolayer
For the first path to transition with cell-cell adhesion as the control parameter, $\bar\alpha=1.25$ and for the second path with active traction force as the control parameter, $\bar\omega=0.5$.

\subsection*{Dynamics of cell polarity}

The dynamics of cell polarity is introduced based on contact inhibition of locomotion~\citep{Abercrombie_1954,Abercrombie_1979} by aligning the polarity of the cell to the direction of the total interaction force acting on the cell \citep{Smeets2016,Peyret2019}. As such, the polarization dynamics is given by: 
\begin{eqnarray}
\label{eq4}
\partial_{t}\theta_{i}=-\frac{1}{\tau}|\vec{T}_{i}|\Delta\theta_{i}+D_{r}\eta,
\end{eqnarray}
where $\theta_{i}\in[-\pi,\pi]$ is the counterclockwise angle of cell polarity measured from $\vec{e}_{x}$, $\hat{F}^{\text{pol}}_{i}=\left(\cos\theta_{i},\sin\theta_{i}\right)$ and $\eta$ is the Gaussian white noise with zero mean, unit variance, $D_{r}$ is rotational diffusivity, $\Delta\theta_{i}$ is the angle between $\hat{F}^{\text{pol}}_{i}$ and $\vec{T}_{i}$, and positive constant $\tau$ sets the alignment time scale.

\subsection*{Acknowledgments}
S.M. is grateful for the generous support of the Rosenfeld Foundation fellowship at the Niels Bohr Institute, University of Copenhagen. S.M., G.R. and J.A. acknowledge support for this research provided by US ARO funding through the Multidisciplinary University Research Initiative (MURI) Grant No. W911NF-19-1-0245. A. D. acknowledges funding from the Novo Nordisk Foundation (grant No. NNF18SA0035142 and NERD grant No. NNF21OC0068687), Villum Fonden Grant no. 29476, and the European Union via the ERC-Starting Grant PhysCoMeT.
\bibliography{main.bib}
\end{document}